\documentclass[twocolumn,aps,prl,showpacs,superscriptaddress]{revtex4}

\usepackage{amssymb,amsmath}
\usepackage{graphicx}
\usepackage[amssymb,pstricks]{SIunits}
\usepackage{color}

\begin{document}

\title{Large quality factor in sheet metamaterials made from dark dielectric meta-atoms}

\author{Aditya~Jain}
\affiliation{Ames Laboratory---U.S. DOE and Department of Physics and Astronomy, Iowa State University, Ames, Iowa 50011, USA}
\author{Philippe~Tassin}
\altaffiliation[Present address: ]{Department of Applied Physics, Chalmers University, SE-412 96 G\"oteborg, Sweden}
\affiliation{Ames Laboratory---U.S. DOE and Department of Physics and Astronomy, Iowa State University, Ames, Iowa 50011, USA}
\author{Thomas~Koschny}
\affiliation{Ames Laboratory---U.S. DOE and Department of Physics and Astronomy, Iowa State University, Ames, Iowa 50011, USA}
\author{Costas~M.~Soukoulis}
\affiliation{Ames Laboratory---U.S. DOE and Department of Physics and Astronomy, Iowa State University, Ames, Iowa 50011, USA}
\affiliation{Institute of Electronic Structure and Lasers (IESL), FORTH, 71110 Heraklion, Crete, Greece}

\date{\today}

\begin{abstract}
Metamaterials---or artificial electromagnetic materials---can create media with properties unattainable in nature, but mitigating dissipation is a key challenge for their further development. Here we demonstrate a low-loss metamaterial by exploiting dark bound states in dielectric inclusions coupled to the external waves by small nonresonant metallic antennas. We experimentally demonstrate a dispersion-engineered metamaterial based on a meta-atom made from alumina and we show that its resonance has a much larger quality factor than metal-based meta-atoms. Finally, we show that our dielectric meta-atom can be used to create sheet metamaterials with negative permittivity or permeability.
\end{abstract}

\maketitle

\begin{figure*}
\centering
\includegraphics[clip,scale=0.75]{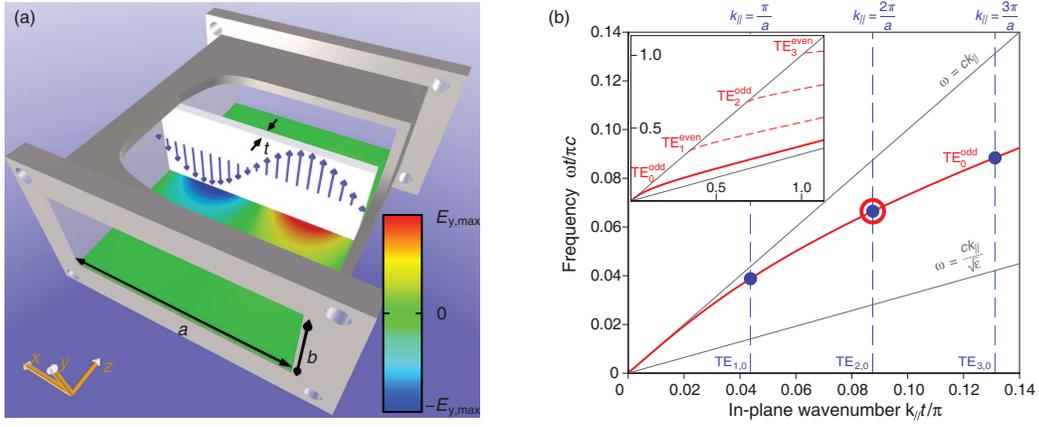}
\caption{Illustration of the dielectric meta-atom and its eigenmode dispersion diagram. (a)~The meta-atom is an alumina slab of thickness $t = \unit{1.016}{\milli\meter}$ (0.04 inch) and its size is $a = \unit{22.86}{\milli\meter}$ and $b = \unit{10.16}{\milli\meter}$. The color plot on the floor of the waveguide and the arrow plot on the slab represent the electric field distribution $E_\mathrm{y}$ of the second quantized state [red circle in Fig.~\ref{Fig1}(b)]. (b)~The red lines constitute the dispersion diagram of the dielectric slab. The vertical blue lines indicate the quantization of the eigenmodes. The blue dots are the quantized eigenstates of the dielectric meta-atom shown in (a) and we work with the second quantized state (red circle) throughout this work. The inset is a zoom-out showing higher-order modes.}\label{Fig1}
\end{figure*}

Metamaterials (MM), i.e., artificial media in which small electromagnetic scatterers replace atoms as the basic elements of interaction with radiation, have led to a myriad of optical properties not attainable or much weaker in natural materials~\cite{Smith-2004,Engheta-2006,Shalaev-2006,Liu-2011,Soukoulis-2011,Zheludev-2012}. Of particular interest are the so-called sheet MMs that consist of a single layer of meta-atoms, because they allow for miniaturized subwavelength-sized microwave and photonic devices for previously unimaginable terahertz and optical wave manipulation~\cite{Zheludev-2012,Engheta-2002,Chen-2006,Engheta-2007,Tassin-2008,Monticone-2013}. Sheet MMs, which have the most promise for major advances in wave technology in the short term, are the subject of this Letter.

Historically, the most popular constitutive elements for electromagnetic MMs turned out to be metallic resonators such as split rings and fishnets with all their variants~\cite{Soukoulis-2011}, although they are plagued by considerable dissipation~\cite{Lagarkov-2010,Boltasseva-2011,Khurgin-2011,Tassin-2012,Hess-2012}. It was established as early as 2003~\cite{OBrien-2002,Holloway-2003} that negative permeability can also be obtained in MMs with dielectric inclusions using Mie resonances~\cite{Zhao-2009}. Many all-dielectric MMs with pronounced resonances have now been demonstrated in the microwave and terahertz bands~\cite{Vendik-2006,Peng-2007,Ueda-2007,Zhao-2008,Popa-2008,Nemec-2012,Wheeler-2005,Schuller-2007,Vynck-2009,Ginn-2012}. The major advantage of dielectric inclusions is that they allow circumventing the Joule heating loss that is detrimental in metallic meta-atoms. However, the size of resonant dielectric particles is typically only slightly smaller than the free-space wavelength~\cite{Soukoulis-2011,Zhao-2009}, which inevitably leads to the breakdown of the effective medium approximation. For purposes of reference, we have performed a simulation of a dielectric square-base cylinder made of alumina. In order to obtain a resonance at \unit{10}{\giga\hertz}, we must increase the thickness of the cylinder to \unit{11}{\milli\meter}, which is only slightly below the free-space wavelength of \unit{30}{\milli\meter}. Fig.~S1~\cite{supplement} plots the retrieved permittivity and permeability of this MM. It has indeed a resonance with a large Q-factor (Q = 263), but we also observe strong deviations from the Lorentzian resonance shape. This is the direct result of operating at the very edge of the effective medium approximation.

The reason for the large size of resonant dielectric particles is the limited polarizability with permittivities typically between 1 and 10 (one can find larger $\epsilon$ in ferroelectric/polaritonic materials, but at the expense of prohibitively high material loss). Therefore, standing waves in dielectric inclusions are only a few times smaller than the free-space wavelength. In this Letter, we present a way to circumvent this problem by using dark (nonradiative) dielectric resonators rather than scattering Mie resonances. This allows us to use one mechanism for the resonance responsible for the stored energy and another mechanism for the coupling of the resonance to the incident wave. We will show that the meta-atom we construct in this way can be used as a versatile building block to create sheet MMs with large quality factors. This is a desirable feature, since it allows properties such as $\epsilon = -1$ or $\mu = -1$ to be achieved at frequencies farther away from the resonance, ultimately reducing dissipation~\cite{Tassin-2012}.

\begin{figure*}
\centering
\includegraphics[clip,scale=0.75]{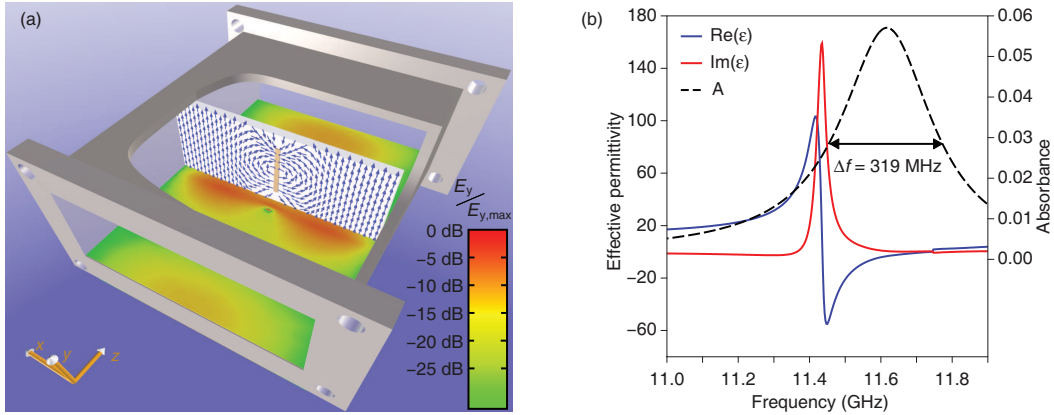}
\caption{Illustration of the cut wire as secondary scatterer. (a)~A cut wire placed on top of the dielectric slab acts as a secondary scatterer to couple the external wave to the nonradiative eigenstate of the dielectric meta-atom. The color plot shows the electric field ($E_\mathrm{y}$) of the electric dipole resonance and the arrows show the dipole fields in the vertical plane. (b)~Effective permittivity and absorbance of a 6.0-mm-long cut wire. The resonance is the electric dipole resonance of the cut wire.}\label{Fig2}
\end{figure*}

Here we use a thin alumina slab as shown in Fig.~\ref{Fig1}(a) (but the resonator can be replaced by other structures with negligible dipole moment) placed in a WR-90 waveguide. The waveguide act as a mirroring condition creating an array of meta-atoms. We utilize the bound states of the thin slab (TE-polarization), of which the dispersion relation is shown in Fig.~\ref{Fig1}(b). Each mode is further quantized by the periodicity of the MM as indicated by the blue disks in Fig.~\ref{Fig1}(b). Since the dispersion relation of the chosen lowest-order mode ($\mathrm{TE}_0^\mathrm{(odd)}$) has no cut-off frequency, the meta-atom can be made arbitrarily thin. We use the second of the quantized states [highlighted with the red circle in Fig.~\ref{Fig1}(b)], of which the electric field, shown in Fig.~\ref{Fig1}(a), is antisymmetric with respect to the center of the slab.

We excite the MM with a normally incident wave. Since the bound state is nonradiative (i.e., its fields are purely evanescent outside the slab), the external wave cannot directly interact with the bound state. Therefore, we have added small antennas acting as secondary scatterers that couple the incident wave to the bound state of the meta-atom. Here we have used metallic cut wires, but the actual shape or material of the secondary scatterers is less important as long as they provide a way to exchange energy between the external wave and the bound state. Figure~\ref{Fig2}(a) plots the electric field distribution of the cut wire. The effective permittivity generated by a 6.0-mm-long cut wire is shown in Fig.~\ref{Fig2}(b). It has a resonance at \unit{11.4}{\giga\hertz} with a quality factor $Q_\mathrm{wire} = 36$. The centered cut wire is of course a conventional metallic meta-atom, so we can use this value as a reference further on. Because of its mirror symmetry, the cut-wire's eigenfields have zero overlap with the antisymmetric bound state of the dielectric slab if the cut wire is centered in the middle of the slab. If we move the cut wire along the $x$-direction, energy exchange can take place in a controlled way.

\begin{figure}
\centering
\includegraphics[clip,scale=0.75]{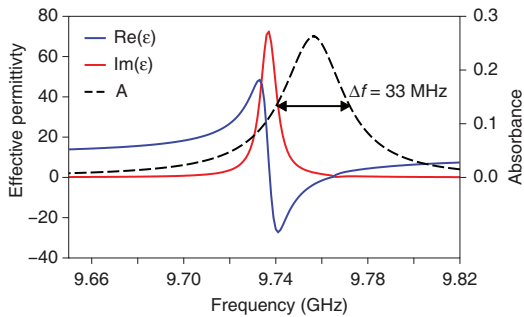}
\caption{Effective permittivity of a MM consisting of the dielectric meta-atom and a 5.0-mm-long wire as secondary scatterer displaced over \unit{2}{\milli\meter} from the center. The high-quality resonance of the dielectric meta-atom generates a negative-permittivity resonance with a bandwidth of  $f = \unit{33}{\mega\hertz}$ and an inverse relative bandwidth of $f_0/\Delta{f} = 296$. This resonance is due to the dark mode in the dielectric slab and not the cut-wire resonance which is at \unit{12.2}{\giga\hertz}.}\label{Fig3}
\end{figure}

First, we have studied the dielectric meta-atom with a cut wire displaced over \unit{2}{\milli\meter} horizontally. The cut wire is now \unit{5}{\milli\meter} long, so that the electric dipole resonance of the wire-on-substrate is at \unit{12.2}{\giga\hertz}, i.e., far away from the frequency of interest (see Fig.~S2~\cite{supplement} for a wideband spectrum from which it is clear that the cut-wire resonance is well separated) and can be ignored for the rest of the discussion. Figure~\ref{Fig3} plots the effective permittivity and the absorbance obtained from simulations (experimental results shown in Fig.~S3 are in good agreement~\cite{supplement}). We observe a very strong Lorentzian-shaped resonance (i.e., with no periodicity artefacts, indicating that the meta-atom is sufficiently subwavelength) with a bandwidth of \unit{33}{\mega\hertz} or an inverse relative bandwidth of $f_0/\Delta{f} = 296$. This resonance is not the cut-wire resonance, but it is the dark mode in the dielectric slab coupled to the incident wave by the nonresonant cut-wire antenna. This result demonstrates that our dielectric meta-atom can produce effective materials with negative permittivity and high-quality resonances. The quality factor is so large because of the large amount of electromagnetic energy that can be stored in the dielectric. Because we are operating far from the wire's own resonance, the current in the wire is out of phase with the electric field, resulting in strongly reduced ohmic losses in the wire.

\begin{figure*}
\centering
\includegraphics[clip]{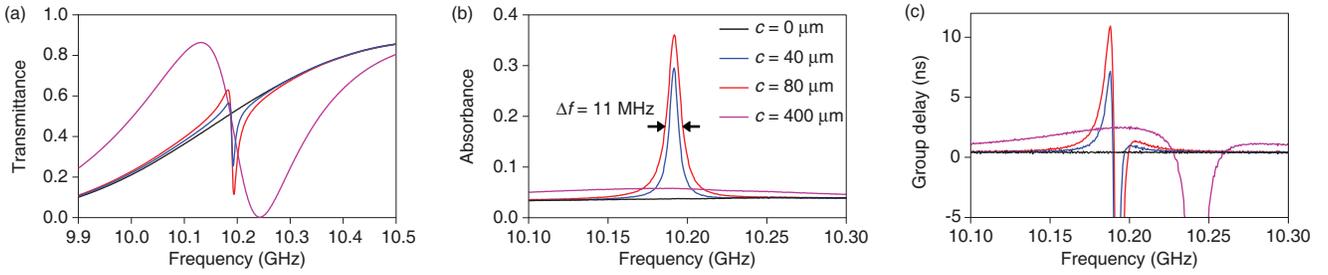}
\caption{Transmission, absorbance and group delay of the MM with the dielectric meta-atom and a 7.79-mm-long cut wire as secondary scattering antenna.  The MM exhibits a very narrow spectral feature indicative of a strong resonance.  (a)-(c), Obtained from experiments. (d)-(f), Obtained from numerical simulations.}\label{Fig4}
\end{figure*}

To confirm the value of the quality factor, we have fabricated the MM, consisting of the dielectric meta-atom and the cut-wire antenna as secondary scatterer~\cite{supplement}. The dielectric resonator was made from a 1.016-mm-thick alumina substrate and the secondary scatterer from copper (Cu). To facilitate the determination of the quality factor of the nonradiative eigenmode, both resonators were now designed to have their resonances at the same frequency (\unit{10.19}{\giga\hertz}). Figures~\ref{Fig4}(a)-(c) plot the resulting transmittance, absorbance and group delay spectra for several values of the coupling strength between the dielectric meta-atom and the secondary scatterer. When the cut wire is centered ($c = 0$, black lines), the bound state in the dielectric slab cannot be excited. The flat (slowly-increasing) background in the transmittance spectra is due to the cut-wire antenna. When we gradually increase the coupling $c$, a narrow spectral feature emerges at \unit{10.19}{\giga\hertz}. In the absorbance, a sharp peak develops and reaches a maximum amplitude for $c = \unit{80}{\micro\meter}$. The inverse relative bandwidth of the absorption feature at $c = \unit{80}{\micro\meter}$ equals $f_0/\Delta{f} = 902$, which is again an indication of the high quality factor of the dielectric resonator (compare with the quality factor $Q_\mathrm{wire} = 36$ of the conventional cut-wire resonator discussed above). We also observe a large group delay of \unit{11}{\nano\second} [see Fig.~\ref{Fig4}(c)]---this amounts to a slowdown of electromagnetic radiation by a factor 3248 with respect to free-space propagation, providing further evidence that electromagnetic energy gets trapped in a high-quality electromagnetic resonator.

\begin{figure}[b!]
\centering
\includegraphics[clip,scale=0.75]{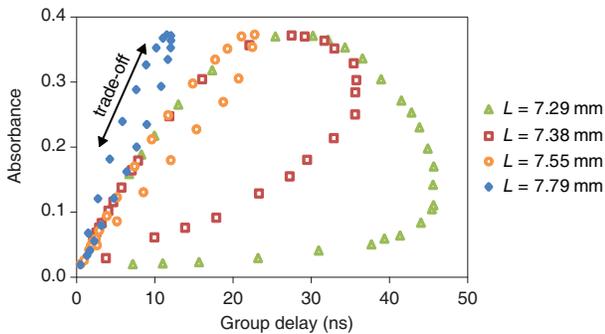}
\caption{Absorbance vs. group delay of the MM when the length of the cut-wire antenna is changed from $L = \unit{7.79}{\milli\meter}$ to $\unit{7.29}{\milli\meter}$ (this effectively moves the resonance frequency of the secondary scatterer to higher frequency).}\label{Fig5}
\end{figure}

The relative linewidth of the spectral feature does not give us directly the quality factor, because of the interaction between the cut wire and the dielectric meta-atom~\cite{Pozar-2005}. However, using an appropriate two-resonator model~\cite{Tassin-2012,Verslegers-2012}, we can extract the quality factor from the transmittance $T$, the group delay $\tau_\mathrm{g}$, and the resonance frequency $f_0$: $Q_\mathrm{exp} = \pi f_0 \tau_\mathrm{g}/[T(1 - T)] = 1728$. This experimental result was verified by an eigenmode simulation of the dielectric meta-atom. The simulation results are shown in Figs.~S4(d)-(f) and excellent agreement with the experimental spectra is obtained. An eigenmode analysis of the dielectric meta-atom without the wire then revealed the quality factor to be $Q_\mathrm{sim} = 1727$. This is significantly higher than what can be obtained with metallic meta-atoms (e.g., the quality factor of the cut wire itself is $Q_\mathrm{wire} \approx 36$). From the simulations, we deduce that the Q-factor is limited in our experiment by (i)~dissipative loss in the alumina slab (loss tangent) and (ii) Ohmic loss due to the finite conductivity of the WR-90 waveguide.

\begin{figure}
\centering
\includegraphics[clip,scale=0.75]{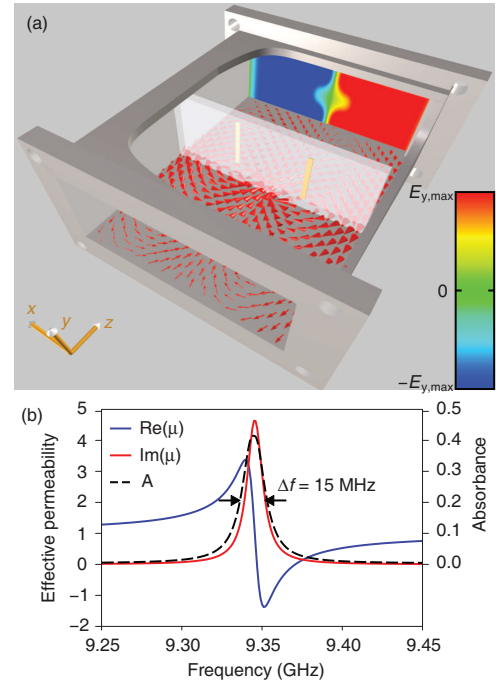}
\caption{(a) A magnetic MM obtained by using a 5.5-mm-long cut-wire pair as secondary scatterer.  The red arrows plot the magnetic field showing that the currents in the wires flow in opposite directions and the surface plot on the back plots the electric field ($E_\mathrm{y}$) in the center of the slab showing that the bound state is excited. (b) Effective permeability.}\label{Fig6}
\end{figure}

A change in interaction strength between the dielectric meta-atom and the secondary scatterer results in a trade-off between the group delay and absorption level (see the curve $L = \unit{7.79}{\milli\meter}$ in Fig.~\ref{Fig5}). Until now, we have chosen the maximum-absorption frequency of the cut wire equal to the maximum-absorption frequency of the dielectric meta-atom, but this is not necessary. We have determined the absorbance for cut wires with different lengths (effectively moving the resonance frequency of the cut-wire antenna) and we find that the group delay can be further increased by reducing the length of the cut wire. For example, with a 7.29-mm-long wire antenna, we can achieve a group delay of \unit{28}{\nano\second} at an equal absorption level of 38\%. Experimental results for the MM with the shortest cut wire, provided in Fig.~S5~\cite{supplement}, confirm these findings. This last result is very interesting, because it clearly demonstrates that the secondary scatterer needs only to provide a coupling mechanism between the dielectric meta-atom and the external waves, but it need not be resonant itself. This can be used to our advantage because it provides a way to reduce resonant losses in the (metallic) secondary scatterer.

Since the bound state of the dielectric meta-atom is nonradiative, it cannot directly create large dipole moments that can reradiate into propagating waves. However, we can add nanostructures with a dipole moment as the secondary scatterers, similar to how negative permeability can be achieved by coupling to surface plasmon polaritons on a metal interface~\cite{Guney-2011}. In Fig.~\ref{Fig6}(a), we show another MM in which the dielectric meta-atom is now decorated by a cut-wire pair across the dielectric slab. The wire pair itself has a magnetic resonance at \unit{11.6}{\giga\hertz}, i.e., it is far away from the dark mode. When the cut-wire pair is coupled to the bound state, we can achieve a magnetic dipole while at the same time avoiding the large resonant loss of the metallic structure. The retrieved permeability shown in Fig.~\ref{Fig6}(b) demonstrates that we can indeed achieve negative permeability in this way. We observe a very sharp resonance indicating a very large quality factor ($f_0/\Delta{f} = 623$). Our experiments (results shown in Fig.~S7) again confirm these findings~\cite{supplement}. We emphasize once more that this resonance is the dark mode in the dielectric inclusion.

Our study shows that bound states in dielectric structures can be used as constituents of resonant MMs by coupling them to the incident waves by small nonresonant antennas. This mechanism provides for much larger quality factors than metallic meta-atoms. Because the dielectric particle does not need a dipole moment of its own, the meta-atoms can be made very thin. As a result, the sheet MMs we have presented here are perfectly homogenizable (no visible periodicity artefacts in the effective constitutive parameters), unlike Mie-resonance MMs that suffer from strong periodicity effects (compare the perfectly Lorentzian shapes in Figs.~3, 4 and 6 with the obvious deviations from the Lorentzian shape in Fig.~S1).

The design strategy demonstrated here is also very versatile. Here we have shown a slow-light MM and a negative-$\mu$ MM, but one could likewise design, e.g., a MM with strong dichroism or optical activity by using a chiral scatterer or MMs with nonlinear functionality by using a nonlinear scatterer. In this way, the dielectric meta-atom offers a universal design framework in which we can modify the response of the MM by a well-chosen nonresonant secondary scatterer, while at the same time the response is dramatically enhanced by the strong, low-loss resonance of the dielectric meta-atom. The proposed dielectric MMs will also be interesting for, a.o., wavefront-engineering metasurfaces and dark-mode sensing.

Work at Ames Lab was partially supported by the U.S.\ Dept.\ of Energy, Basic Energy Science, Materials Sciences and Engineering, Contract No.\ DE-AC02-07CH11358 (experiments), and by the U.S.\ Office of Naval Research, Award No.\ N00014-10-1-0925 (theory).


\begin{thebibliography}{32}
\expandafter\ifx\csname natexlab\endcsname\relax\def\natexlab#1{#1}\fi
\expandafter\ifx\csname bibnamefont\endcsname\relax
  \def\bibnamefont#1{#1}\fi
\expandafter\ifx\csname bibfnamefont\endcsname\relax
  \def\bibfnamefont#1{#1}\fi
\expandafter\ifx\csname citenamefont\endcsname\relax
  \def\citenamefont#1{#1}\fi
\expandafter\ifx\csname url\endcsname\relax
  \def\url#1{\texttt{#1}}\fi
\expandafter\ifx\csname urlprefix\endcsname\relax\def\urlprefix{URL }\fi
\providecommand{\bibinfo}[2]{#2}
\providecommand{\eprint}[2][]{\url{#2}}

\bibitem[{\citenamefont{Smith et~al.}(2004)\citenamefont{Smith, Pendry, and
  Wiltshire}}]{Smith-2004}
\bibinfo{author}{\bibfnamefont{D.~R.} \bibnamefont{Smith}},
  \bibinfo{author}{\bibfnamefont{J.~B.} \bibnamefont{Pendry}},
  \bibnamefont{and} \bibinfo{author}{\bibfnamefont{M.~C.~K.}
  \bibnamefont{Wiltshire}}, \bibinfo{journal}{Science}
  \textbf{\bibinfo{volume}{305}}, \bibinfo{pages}{788} (\bibinfo{year}{2004}).

\bibitem[{\citenamefont{Engheta and Ziolkowski}(2006)}]{Engheta-2006}
\bibinfo{author}{\bibfnamefont{R.}~\bibnamefont{Engheta}} \bibnamefont{and}
  \bibinfo{author}{\bibfnamefont{R.~W.} \bibnamefont{Ziolkowski}},
  \emph{\bibinfo{title}{Metamaterials, Physics and Engineering Explorations}}
  (\bibinfo{publisher}{Wiley-IEEE Press}, \bibinfo{address}{New York},
  \bibinfo{year}{2006}).

\bibitem[{\citenamefont{Shalaev}(2006)}]{Shalaev-2006}
\bibinfo{author}{\bibfnamefont{V.~M.} \bibnamefont{Shalaev}},
  \bibinfo{journal}{Nature Photon.} \textbf{\bibinfo{volume}{1}},
  \bibinfo{pages}{41} (\bibinfo{year}{2006}).

\bibitem[{\citenamefont{Liu and Zhang}(2011)}]{Liu-2011}
\bibinfo{author}{\bibfnamefont{Y.}~\bibnamefont{Liu}} \bibnamefont{and}
  \bibinfo{author}{\bibfnamefont{X.}~\bibnamefont{Zhang}},
  \bibinfo{journal}{Chem. Soc. Rev.} \textbf{\bibinfo{volume}{40}},
  \bibinfo{pages}{2494} (\bibinfo{year}{2011}).

\bibitem[{\citenamefont{Soukoulis and Wegener}(2011)}]{Soukoulis-2011}
\bibinfo{author}{\bibfnamefont{C.~M.} \bibnamefont{Soukoulis}}
  \bibnamefont{and} \bibinfo{author}{\bibfnamefont{M.}~\bibnamefont{Wegener}},
  \bibinfo{journal}{Nature Photon.} \textbf{\bibinfo{volume}{5}},
  \bibinfo{pages}{523} (\bibinfo{year}{2011}).

\bibitem[{\citenamefont{Zheludev and Kivshar}(2012)}]{Zheludev-2012}
\bibinfo{author}{\bibfnamefont{N.~I.} \bibnamefont{Zheludev}} \bibnamefont{and}
  \bibinfo{author}{\bibfnamefont{Y.~S.} \bibnamefont{Kivshar}},
  \bibinfo{journal}{Nature Mater.} \textbf{\bibinfo{volume}{11}},
  \bibinfo{pages}{917} (\bibinfo{year}{2012}).

\bibitem[{\citenamefont{Shelby et~al.}(2002)\citenamefont{Shelby, Smith, and
  Schultz}}]{Engheta-2002}
\bibinfo{author}{\bibfnamefont{R.~A.} \bibnamefont{Shelby}},
  \bibinfo{author}{\bibfnamefont{D.~R.} \bibnamefont{Smith}}, \bibnamefont{and}
  \bibinfo{author}{\bibfnamefont{S.}~\bibnamefont{Schultz}},
  \bibinfo{journal}{IEEE Ant. Wireless Prop. Lett.}
  \textbf{\bibinfo{volume}{1}}, \bibinfo{pages}{10} (\bibinfo{year}{2002}).

\bibitem[{\citenamefont{Chen et~al.}(2006)\citenamefont{Chen, Padilla, Zide,
  Gossard, Taylor, and Averitt}}]{Chen-2006}
\bibinfo{author}{\bibfnamefont{H.-T.} \bibnamefont{Chen}},
  \bibinfo{author}{\bibfnamefont{W.~J.} \bibnamefont{Padilla}},
  \bibinfo{author}{\bibfnamefont{J.~M.~O.} \bibnamefont{Zide}},
  \bibinfo{author}{\bibfnamefont{A.~C.} \bibnamefont{Gossard}},
  \bibinfo{author}{\bibfnamefont{A.~J.} \bibnamefont{Taylor}},
  \bibnamefont{and} \bibinfo{author}{\bibfnamefont{R.~D.}
  \bibnamefont{Averitt}}, \bibinfo{journal}{Nature}
  \textbf{\bibinfo{volume}{444}}, \bibinfo{pages}{597} (\bibinfo{year}{2006}).

\bibitem[{\citenamefont{Engheta}(2007)}]{Engheta-2007}
\bibinfo{author}{\bibfnamefont{N.}~\bibnamefont{Engheta}},
  \bibinfo{journal}{Science} \textbf{\bibinfo{volume}{317}},
  \bibinfo{pages}{1698} (\bibinfo{year}{2007}).

\bibitem[{\citenamefont{Tassin et~al.}(2008)\citenamefont{Tassin, Sahyoun, and
  Veretennicoff}}]{Tassin-2008}
\bibinfo{author}{\bibfnamefont{P.}~\bibnamefont{Tassin}},
  \bibinfo{author}{\bibfnamefont{X.}~\bibnamefont{Sahyoun}}, \bibnamefont{and}
  \bibinfo{author}{\bibfnamefont{I.}~\bibnamefont{Veretennicoff}},
  \bibinfo{journal}{Appl.\ Phys.\ Lett.} \textbf{\bibinfo{volume}{92}},
  \bibinfo{pages}{203111} (\bibinfo{year}{2008}).

\bibitem[{\citenamefont{Monticone et~al.}(2013)\citenamefont{Monticone,
  Estakhri, and Alu}}]{Monticone-2013}
\bibinfo{author}{\bibfnamefont{F.}~\bibnamefont{Monticone}},
  \bibinfo{author}{\bibfnamefont{N.~M.} \bibnamefont{Estakhri}},
  \bibnamefont{and} \bibinfo{author}{\bibfnamefont{A.}~\bibnamefont{Alu}},
  \bibinfo{journal}{Phys.\ Rev.\ Lett.} \textbf{\bibinfo{volume}{110}},
  \bibinfo{pages}{203903} (\bibinfo{year}{2013}).

\bibitem[{\citenamefont{Lagarkov et~al.}(2010)\citenamefont{Lagarkov, Kisel,
  and Sarychev}}]{Lagarkov-2010}
\bibinfo{author}{\bibfnamefont{A.~N.} \bibnamefont{Lagarkov}},
  \bibinfo{author}{\bibfnamefont{V.~N.} \bibnamefont{Kisel}}, \bibnamefont{and}
  \bibinfo{author}{\bibfnamefont{A.~K.} \bibnamefont{Sarychev}},
  \bibinfo{journal}{J.\ Opt.\ Soc.\ Am.\ B} \textbf{\bibinfo{volume}{27}},
  \bibinfo{pages}{648} (\bibinfo{year}{2010}).

\bibitem[{\citenamefont{Boltasseva and Atwater}(2011)}]{Boltasseva-2011}
\bibinfo{author}{\bibfnamefont{A.}~\bibnamefont{Boltasseva}} \bibnamefont{and}
  \bibinfo{author}{\bibfnamefont{H.~A.} \bibnamefont{Atwater}},
  \bibinfo{journal}{Science} \textbf{\bibinfo{volume}{331}},
  \bibinfo{pages}{290} (\bibinfo{year}{2011}).

\bibitem[{\citenamefont{Khurgin and Sun}(2011)}]{Khurgin-2011}
\bibinfo{author}{\bibfnamefont{J.~B.} \bibnamefont{Khurgin}} \bibnamefont{and}
  \bibinfo{author}{\bibfnamefont{G.}~\bibnamefont{Sun}},
  \bibinfo{journal}{Appl.\ Phys.\ Lett.} \textbf{\bibinfo{volume}{99}},
  \bibinfo{pages}{211106} (\bibinfo{year}{2011}).

\bibitem[{\citenamefont{{Tassin} et~al.}(2012)\citenamefont{{Tassin},
  {Koschny}, {Kafesaki}, and {Soukoulis}}}]{Tassin-2012}
\bibinfo{author}{\bibfnamefont{P.}~\bibnamefont{{Tassin}}},
  \bibinfo{author}{\bibfnamefont{T.}~\bibnamefont{{Koschny}}},
  \bibinfo{author}{\bibfnamefont{M.}~\bibnamefont{{Kafesaki}}},
  \bibnamefont{and} \bibinfo{author}{\bibfnamefont{C.~M.}
  \bibnamefont{{Soukoulis}}}, \bibinfo{journal}{Nature Photon.}
  \textbf{\bibinfo{volume}{6}}, \bibinfo{pages}{259} (\bibinfo{year}{2012}).

\bibitem[{\citenamefont{Hess et~al.}(2012)\citenamefont{Hess, Pendry, Maier,
  Oulton, Hamm, and Tsakmakidis}}]{Hess-2012}
\bibinfo{author}{\bibfnamefont{O.}~\bibnamefont{Hess}},
  \bibinfo{author}{\bibfnamefont{J.~B.} \bibnamefont{Pendry}},
  \bibinfo{author}{\bibfnamefont{S.~A.} \bibnamefont{Maier}},
  \bibinfo{author}{\bibfnamefont{R.~F.} \bibnamefont{Oulton}},
  \bibinfo{author}{\bibfnamefont{J.~M.} \bibnamefont{Hamm}}, \bibnamefont{and}
  \bibinfo{author}{\bibfnamefont{K.~L.} \bibnamefont{Tsakmakidis}},
  \bibinfo{journal}{Nature Mater.} \textbf{\bibinfo{volume}{11}},
  \bibinfo{pages}{573} (\bibinfo{year}{2012}).

\bibitem[{\citenamefont{O'Brien and Pendry}(2002)}]{OBrien-2002}
\bibinfo{author}{\bibfnamefont{S.}~\bibnamefont{O'Brien}} \bibnamefont{and}
  \bibinfo{author}{\bibfnamefont{J.~B.} \bibnamefont{Pendry}},
  \bibinfo{journal}{J.\ Phys.\ Condens.\ Matter} \textbf{\bibinfo{volume}{14}},
  \bibinfo{pages}{4035} (\bibinfo{year}{2002}).

\bibitem[{\citenamefont{Holloway et~al.}(2003)\citenamefont{Holloway, Kuester,
  Baker-Jarvis, and Kabos}}]{Holloway-2003}
\bibinfo{author}{\bibfnamefont{C.~L.} \bibnamefont{Holloway}},
  \bibinfo{author}{\bibfnamefont{E.~F.} \bibnamefont{Kuester}},
  \bibinfo{author}{\bibfnamefont{J.}~\bibnamefont{Baker-Jarvis}},
  \bibnamefont{and} \bibinfo{author}{\bibfnamefont{P.~A.} \bibnamefont{Kabos}},
  \bibinfo{journal}{IEEE Trans.\ Antennas Propag.}
  \textbf{\bibinfo{volume}{51}}, \bibinfo{pages}{2596} (\bibinfo{year}{2003}).

\bibitem[{\citenamefont{Zhao et~al.}(2009)\citenamefont{Zhao, Zhou, Zhang, and
  Lippens}}]{Zhao-2009}
\bibinfo{author}{\bibfnamefont{Q.}~\bibnamefont{Zhao}},
  \bibinfo{author}{\bibfnamefont{J.}~\bibnamefont{Zhou}},
  \bibinfo{author}{\bibfnamefont{F.}~\bibnamefont{Zhang}}, \bibnamefont{and}
  \bibinfo{author}{\bibfnamefont{D.}~\bibnamefont{Lippens}},
  \bibinfo{journal}{Mater.\ Today} \textbf{\bibinfo{volume}{12}},
  \bibinfo{pages}{60} (\bibinfo{year}{2009}).

\bibitem[{\citenamefont{Vendik et~al.}(2006)\citenamefont{Vendik, Vendik, and
  Gashinova}}]{Vendik-2006}
\bibinfo{author}{\bibfnamefont{I.~B.} \bibnamefont{Vendik}},
  \bibinfo{author}{\bibfnamefont{O.~G.} \bibnamefont{Vendik}},
  \bibnamefont{and} \bibinfo{author}{\bibfnamefont{M.~S.}
  \bibnamefont{Gashinova}}, \bibinfo{journal}{Techn.\ Phys.\ Lett.}
  \textbf{\bibinfo{volume}{32}}, \bibinfo{pages}{429} (\bibinfo{year}{2006}).

\bibitem[{\citenamefont{Peng et~al.}(2007)\citenamefont{Peng, Ran, Chen, Zhang,
  Kong, and Grzegorczyk}}]{Peng-2007}
\bibinfo{author}{\bibfnamefont{L.}~\bibnamefont{Peng}},
  \bibinfo{author}{\bibfnamefont{L.}~\bibnamefont{Ran}},
  \bibinfo{author}{\bibfnamefont{H.}~\bibnamefont{Chen}},
  \bibinfo{author}{\bibfnamefont{H.}~\bibnamefont{Zhang}},
  \bibinfo{author}{\bibfnamefont{J.~A.} \bibnamefont{Kong}}, \bibnamefont{and}
  \bibinfo{author}{\bibfnamefont{T.~M.} \bibnamefont{Grzegorczyk}},
  \bibinfo{journal}{Phys.\ Rev.\ Lett.} \textbf{\bibinfo{volume}{98}},
  \bibinfo{pages}{157403} (\bibinfo{year}{2007}).

\bibitem[{\citenamefont{Ueda et~al.}(2007)\citenamefont{Ueda, Lai, and
  Itoh}}]{Ueda-2007}
\bibinfo{author}{\bibfnamefont{T.}~\bibnamefont{Ueda}},
  \bibinfo{author}{\bibfnamefont{A.}~\bibnamefont{Lai}}, \bibnamefont{and}
  \bibinfo{author}{\bibfnamefont{T.}~\bibnamefont{Itoh}},
  \bibinfo{journal}{IEEE Trans.\ Microw.\ Theory Tech.}
  \textbf{\bibinfo{volume}{55}}, \bibinfo{pages}{1280} (\bibinfo{year}{2007}).

\bibitem[{\citenamefont{Zhao et~al.}(2008)\citenamefont{Zhao, Kang, Du, Zhao,
  Xie, Huang, Li, Zhou, and Li}}]{Zhao-2008}
\bibinfo{author}{\bibfnamefont{Q.}~\bibnamefont{Zhao}},
  \bibinfo{author}{\bibfnamefont{L.}~\bibnamefont{Kang}},
  \bibinfo{author}{\bibfnamefont{B.}~\bibnamefont{Du}},
  \bibinfo{author}{\bibfnamefont{H.}~\bibnamefont{Zhao}},
  \bibinfo{author}{\bibfnamefont{Q.}~\bibnamefont{Xie}},
  \bibnamefont{et~al.},
  \bibinfo{journal}{Phys.\ Rev.\ Lett.} \textbf{\bibinfo{volume}{101}},
  \bibinfo{pages}{027402} (\bibinfo{year}{2008}).

\bibitem[{\citenamefont{Popa and Cummer}(2008)}]{Popa-2008}
\bibinfo{author}{\bibfnamefont{B.-I.} \bibnamefont{Popa}} \bibnamefont{and}
  \bibinfo{author}{\bibfnamefont{S.~A.} \bibnamefont{Cummer}},
  \bibinfo{journal}{Phys.\ Rev.\ Lett.} \textbf{\bibinfo{volume}{100}},
  \bibinfo{pages}{207401} (\bibinfo{year}{2008}).

\bibitem[{\citenamefont{Nemec et~al.}(2012)\citenamefont{Nemec, Kadlec, Kadlec,
  Kuzel, Yahiaoui, Chung, Elissalde, Maglione, and Mounaix}}]{Nemec-2012}
\bibinfo{author}{\bibfnamefont{H.}~\bibnamefont{Nemec}},
  \bibinfo{author}{\bibfnamefont{C.}~\bibnamefont{Kadlec}},
  \bibinfo{author}{\bibfnamefont{F.}~\bibnamefont{Kadlec}},
  \bibinfo{author}{\bibfnamefont{P.}~\bibnamefont{Kuzel}},
  \bibinfo{author}{\bibfnamefont{R.}~\bibnamefont{Yahiaoui}},
  \bibinfo{author}{\bibfnamefont{U.-C.} \bibnamefont{Chung}},
  \bibinfo{author}{\bibfnamefont{C.}~\bibnamefont{Elissalde}},
  \bibinfo{author}{\bibfnamefont{M.}~\bibnamefont{Maglione}}, \bibnamefont{and}
  \bibinfo{author}{\bibfnamefont{P.}~\bibnamefont{Mounaix}},
  \bibinfo{journal}{Appl.\ Phys.\ Lett.} \textbf{\bibinfo{volume}{100}},
  \bibinfo{pages}{061117} (\bibinfo{year}{2012}).

\bibitem[{\citenamefont{Wheeler et~al.}(2005)\citenamefont{Wheeler, Aitchison,
  and Mojahedi}}]{Wheeler-2005}
\bibinfo{author}{\bibfnamefont{M.~S.} \bibnamefont{Wheeler}},
  \bibinfo{author}{\bibfnamefont{J.~S.} \bibnamefont{Aitchison}},
  \bibnamefont{and} \bibinfo{author}{\bibfnamefont{M.}~\bibnamefont{Mojahedi}},
  \bibinfo{journal}{Phys.\ Rev.\ B} \textbf{\bibinfo{volume}{72}},
  \bibinfo{pages}{193103} (\bibinfo{year}{2005}).

\bibitem[{\citenamefont{Schuller et~al.}(2007)\citenamefont{Schuller, Zia,
  Taubner, and Brongersma}}]{Schuller-2007}
\bibinfo{author}{\bibfnamefont{J.~A.} \bibnamefont{Schuller}},
  \bibinfo{author}{\bibfnamefont{R.}~\bibnamefont{Zia}},
  \bibinfo{author}{\bibfnamefont{T.}~\bibnamefont{Taubner}}, \bibnamefont{and}
  \bibinfo{author}{\bibfnamefont{M.~L.} \bibnamefont{Brongersma}},
  \bibinfo{journal}{Phys.\ Rev.\ Lett.} \textbf{\bibinfo{volume}{99}},
  \bibinfo{pages}{107401} (\bibinfo{year}{2007}).

\bibitem[{\citenamefont{Vynck et~al.}(2009)\citenamefont{Vynck, Felbacq,
  Centeno, Cabuz, Cassagne, and Guizal}}]{Vynck-2009}
\bibinfo{author}{\bibfnamefont{K.}~\bibnamefont{Vynck}},
  \bibinfo{author}{\bibfnamefont{D.}~\bibnamefont{Felbacq}},
  \bibinfo{author}{\bibfnamefont{E.}~\bibnamefont{Centeno}},
  \bibinfo{author}{\bibfnamefont{A.~I.} \bibnamefont{Cabuz}},
  \bibinfo{author}{\bibfnamefont{D.}~\bibnamefont{Cassagne}}, \bibnamefont{and}
  \bibinfo{author}{\bibfnamefont{B.}~\bibnamefont{Guizal}},
  \bibinfo{journal}{Phys.\ Rev.\ Lett.} \textbf{\bibinfo{volume}{102}},
  \bibinfo{pages}{133901} (\bibinfo{year}{2009}).

\bibitem[{\citenamefont{Ginn et~al.}(2012)\citenamefont{Ginn, Brener, Peters,
  Wendt, Stevens, Hines, Basilio, Warne, Ihlefeld, Clem et~al.}}]{Ginn-2012}
\bibinfo{author}{\bibfnamefont{J.~C.} \bibnamefont{Ginn}},
  \bibinfo{author}{\bibfnamefont{I.}~\bibnamefont{Brener}},
  \bibinfo{author}{\bibfnamefont{D.~W.} \bibnamefont{Peters}},
  \bibinfo{author}{\bibfnamefont{J.~R.} \bibnamefont{Wendt}},
  \bibinfo{author}{\bibfnamefont{J.~O.} \bibnamefont{Stevens}},
  \bibinfo{author}{\bibfnamefont{P.~F.} \bibnamefont{Hines}},
  \bibnamefont{et~al.}, \bibinfo{journal}{Phys.\ Rev.\ Lett.}
  \textbf{\bibinfo{volume}{108}}, \bibinfo{pages}{097402}
  (\bibinfo{year}{2012}).

\bibitem[{sup()}]{supplement}
\emph{\bibinfo{title}{See supplemental material for the methods and for additional supporting experimental and
  numerical results.}}

\bibitem[{\citenamefont{Verslegers et~al.}(2012)\citenamefont{Verslegers, Yu,
  Ruan, Catrysse, and Fan}}]{Verslegers-2012}
\bibinfo{author}{\bibfnamefont{L.}~\bibnamefont{Verslegers}},
  \bibinfo{author}{\bibfnamefont{Z.}~\bibnamefont{Yu}},
  \bibinfo{author}{\bibfnamefont{Z.}~\bibnamefont{Ruan}},
  \bibinfo{author}{\bibfnamefont{P.~B.} \bibnamefont{Catrysse}},
  \bibnamefont{and} \bibinfo{author}{\bibfnamefont{S.}~\bibnamefont{Fan}},
  \bibinfo{journal}{Phys.\ Rev.\ Lett.} \textbf{\bibinfo{volume}{108}},
  \bibinfo{pages}{083902} (\bibinfo{year}{2012}).

\bibitem[{\citenamefont{Pozar}(2005)}]{Pozar-2005}
\bibinfo{author}{\bibfnamefont{D.~M.}~\bibnamefont{Pozar}},
  \emph{\bibinfo{title}{Microwave Engineering, 3rd Ed.}}
  (\bibinfo{publisher}{Wiley}, \bibinfo{address}{New York},
  \bibinfo{year}{2005}).

\bibitem[{\citenamefont{G\"uney et~al.}(2011)\citenamefont{G\"uney, Koschny,
  and Soukoulis}}]{Guney-2011}
\bibinfo{author}{\bibfnamefont{D.~O.} \bibnamefont{G\"uney}},
  \bibinfo{author}{\bibfnamefont{T.}~\bibnamefont{Koschny}}, \bibnamefont{and}
  \bibinfo{author}{\bibfnamefont{C.~M.} \bibnamefont{Soukoulis}},
  \bibinfo{journal}{Phys.\ Rev.\ B} \textbf{\bibinfo{volume}{83}},
  \bibinfo{pages}{045107} (\bibinfo{year}{2011}).

\end{thebibliography}

\end{document}